\documentclass{article}
\usepackage{spconf,amsmath,graphicx}
\usepackage{booktabs} % For better table formatting
\usepackage[table]{xcolor} % For cell coloring
\usepackage{enumitem}

\usepackage[colorlinks=true, linkcolor=black, urlcolor=black, citecolor=black]{hyperref}
\usepackage{siunitx}
\usepackage{makecell} % Allows for line breaks within table cells
\setlist{nosep, leftmargin=14pt}

\usepackage{mwe} % to get dummy images

\title{Characterisation of Anti-Arrhythmic Drug Effects on Cardiac Electrophysiology using Physics-Informed Neural Networks}
\name{Ching-En Chiu$^{1}$ \quad Arieh Levy Pinto\sthanks{The second author performed the work
while at Department of Biomedical Engineering, Imperial College London, London, United Kingdom.} \quad Rasheda A Chowdhury$^{2}$\quad Kim Christensen$^{3}$ \quad Marta Varela$^{2}$}

\address{$^{1}$ Department of Electrical and Electronic Engineering, Imperial College London,	United Kingdom\\
$^{2}$ National Heart and Lung Institute, Imperial College London, London, United Kingdom \\ 
$^{3}$ Blackett Laboratory, Imperial College London, London, United Kingdom}
\begin{document}
%\ninept
%
\maketitle
\begin{abstract}
The ability to accurately infer cardiac electrophysiological (EP) properties is key to improving arrhythmia diagnosis and treatment. In this work, we developed a physics-informed neural networks (PINNs) framework to predict how different myocardial EP parameters are modulated by anti-arrhythmic drugs. Using \textit{in vitro} optical mapping images and the 3-channel Fenton-Karma model, we estimated the changes in ionic channel conductance caused by these drugs.

Our framework successfully characterised the action of drugs HMR1556, nifedipine and lidocaine -- respectively, blockade of $I$\textsubscript{K}, $I$\textsubscript{Ca}, and $I$\textsubscript{Na} currents -- by estimating that they decreased the respective channel conductance by $31.8\pm2.7\%$ $(p=8.2 \times 10^{-5})$, $80.9\pm21.6\%$ $(p=0.02)$, and $8.6\pm0.5\%$ $ (p=0.03)$, leaving the conductance of other channels unchanged. For carbenoxolone, whose main action is the blockade of intercellular gap junctions, PINNs also successfully predicted no significant changes $(p>0.09)$ in all ionic conductances.

Our results are an important step towards the deployment of PINNs for model parameter estimation from experimental data, bringing this framework closer to clinical or laboratory images analysis and for the personalisation of mathematical models.

% Physics-Informed Neural Networks (PINNs) are fast becoming an important tool to solve differential equations rapidly
% and accurately, and to identify the systems parameters that
% best agree with a given set of measurements. The latter application is particularly important in biomedical applications, to
% enable the personalisation of mathematical models based on
% the existing clinical or laboratory data. PINNs-enabled model
% parameterisation has been used for cardiac electrophysiology
% (EP), but only for simple EP models and mostly with in silico
% data.
% Here, we demonstrate how this approach can be used on in
% vitro optical mapping images to accurately estimate EP parameters related to ionic channels, which allows us to charaterise the effects of anti-arrhythmic drugs. Our framework
% identified the effects of HMR1556 and Nifedipine - blockade of IKr and ICaL currents - by estimating that they decrease the respective channel conductance by 31.8 ± 2.7%,
% and 80.9 ± 21.6%. Our results are an important

% The abstract should appear at the top of the left-hand column of text, about
% 0.5 inch (12 mm) below the title area and no more than 3.125 inches (80 mm) in
% length.  Leave a 0.5 inch (12 mm) space between the end of the abstract and the
% beginning of the main text.  The abstract should contain about 100 to 150
% words, and should be identical to the abstract text submitted electronically
% along with the paper cover sheet.  All manuscripts must be in English, printed
% in black ink.
\end{abstract}
\section{Introduction}
\label{sec:intro}

% These guidelines include complete descriptions of the fonts, spacing, and
% related information for producing your proceedings manuscripts.
Physics-Informed Neural Networks (PINNs) are a machine learning method that combines knowledge of the equations of a system with data learning \cite{Raissi2019}. This domain knowledge allows PINNs to learn from a small fraction of the data that conventional neural networks require and ensures predictions consistent with physics models. This makes PINNs a suitable candidate for biomedical applications, where data are often sparse and noisy and interpretability is highly desired.

Within cardiac EP, Sahli-Costabal \textit{et al.} used PINNs to solve the isotropic diffusion eikonal equation, using \textit{in silico} data. They estimated arrival times of the action potential (AP) and conduction velocity maps across the left atrial surface~\cite{sahli2020physics}. Following this work, Grandits \textit{et al.} solved the anisotropic eikonal equation to learn fibre orientations and conductivity tensors, using synthetic data and data from one patient~\cite{grandits2021learning}. Closely related to our work, Herrero Martin \textit{et al.} used PINNs with the monodomain equation on sparse maps of transmembrane potential to estimate EP parameters, such as the isotropic diffusion coefficient or surrogates of the AP duration~\cite{herrero2022ep}. The study relied on the Aliev Panfilov model, a simple 2-variable model without clear biological counterparts~\cite{aliev1996simple}. To provide clinically useful characterisation of cardiac EP properties under the influence of drugs, PINNs will need to be deployed in EP models that better capture ionic channel complexity and demonstrate its ability to make inferences from experimental data.

To this end, we coupled PINNs with the Fenton-Karma model, which is a 3-variable model of the cardiac AP with formulations of three transmembrane currents: a fast inward $I_\text{fi}$, slow outward $I_\text{so}$, and slow inward $I_\text{si}$~\cite{fenton1998vortex}. They are analogous to Na$^+$, K$^+$, and Ca$^{2+}$ currents, respectively. We selected this model for its ability to reproduce restitution properties and capture arrhythmic conditions, with the appropriate level of complexity for PINNs.
% The model quantitatively reproduces restitution properties with minimal ionic complexity. 
% PINNs-enabled model parameterisation has been used for cardiac electrophysiology (EP), but only for simple EP models and mostly with \textit{in silico} data. 
% While existing PINNs work mostly placed their focus on \textit{in silico} studies, this work focuses on developing a PINNs framework for the analysis of experimental data.
The Fenton-Karma equations are as follows:
\begin{subequations}
\begin{equation}
\label{FK1} \partial_t u = \nabla \cdot (D\nabla u) - J_\text{fi}(u;v) - J_\text{so}(u) - J_\text{si}(u;w),
\end{equation}
\vspace*{-0.25cm}
\begin{equation}
\partial_t v = \Theta(u_c - u)(1 - v)/\tau_v^{-}(u) - \Theta(u - u_c)v/\tau_v^{+}, \label{FK2}
\end{equation}
\vspace*{-0.25cm}
\begin{equation}
\partial_t w = \Theta(u_c - u)(1 - w)/\tau_w^{-} - \Theta(u - u_c)w/\tau_w^{+}, \label{FK3}
\end{equation}
\end{subequations}where the three scaled currents are given by
\begin{subequations}
\begin{equation}\label{eq:Jfi}
J_\text{fi}(u,v) = -\frac{v}{\tau_d} \Theta(u - u_c)(1 - u)(u - u_c),
\end{equation}
\vspace*{-0.25cm}
\begin{equation}\label{eq:Jso}
J_\text{so}(u) = \frac{u}{\tau_o} \Theta(u_c - u) + \frac{1}{\tau_r} \Theta(u - u_c),
\end{equation}
\vspace*{-0.25cm}
\begin{equation}\label{eq:Jsi}
J_\text{si}(u;w) = -\frac{w}{2\tau_\text{si}} (1 + \tanh[k(u - u_c^\text{si})]).
\end{equation}
\end{subequations}
Here, $u$ is the dimensionless electrical potential across the cell membrane, scaled to take values from $0$ to $1$. $v$ and $w$ are the latent gate variables for $I_\text{fi}$ and $I_\text{si}$, respectively. $\Theta(x)$ is the Heaviside step function defined by $\Theta(x)=1$ for $x\geq0$ and $\Theta(x)=0$ for $x<0$. Membrane potential diffuses to the neighbouring cells, as modelled by the diffusion term for $u$, $\nabla (D\nabla u)$. We assume homogeneous and isotropic conduction, in which case $D$ is a constant scalar. The 8 $\tau$ parameters are various time constants. Among them, $\tau_d$, $\tau_r$, $\tau_o$, and $\tau_{si}$ are inversely proportional to ionic channel conductances, as shown in Eq.~\eqref{eq:Jfi}-\eqref{eq:Jsi}. We model the effects of anti-arrhythmic drugs as time-independent reductions in channel conductances, that is, as increases in the $\tau$ variables.

\textbf{Aims.} We aim to demonstrate how PINNs can characterise the effects of anti-arrhythmic drugs on ionic channels using \textit{in vitro} optical mappings images from cardiomyocyte preparations. We will use PINNs to solve the Fenton-Karma equations and simultaneously estimate model parameters related to ionic channel conductances. For this we will use:
\begin{enumerate}
    \item \textit{in silico} data, which allow us to assess the accuracy and tune the model.
    \item \textit{in vitro} optical mapping data from cardiomyocyte preparations.
\end{enumerate}

\section{Methods}
\label{sec:format}

% All printed material, including text, illustrations, and charts, must be kept
% within a print area of 7 inches (178 mm) wide by 9 inches (229 mm) high. Do
% not write or print anything outside the print area. The top margin must be 1
% inch (25 mm), except for the title page, and the left margin must be 0.75 inch
% (19 mm).  All {\it text} must be in a two-column format. Columns are to be 3.39
% inches (86 mm) wide, with a 0.24 inch (6 mm) space between them. Text must be
% fully justified.

All code used in this study is available at \url{github.com/annien094/EP-PINNs-for-drugs}.

\subsection{\textit{In Silico} Data Generation}
We generated \textit{in silico} membrane potentials on a 1D cable of \num{10}\unit{mm} by solving the Fenton-Karma model with a central finite difference and an explicit 4-stage Runge-Kutta method, using a time step of \num{5}\unit{\micro\second} and a spatial step of \num{100}\unit{\micro\metre}. The values of model parameters were set as in \cite{goodman2005membrane} (see FK-CAF model). We used APs across time from two spatial points as inputs to the PINNs, to mimic the available experimental data. We assessed the accuracy of parameter estimation using the relative error (RE).% and the .
\subsection{Optical Mapping Images}
%- drugs and their effect, what we expect to see
%- what optical mapping is (?) how it is obtained - by electrodes
%- how we process the data

We tested the performance of our PINNs setup on \textit{in vitro} optical mapping images from neonatal rat ventricular myocyte preparations. These preparations were stained with a voltage-sensitive dye (di-8-ANNEPS) before imaging at a high spatio-temporal resolution of \num{400}$\times$\num{85} pixels at \num{525.39} frames/sec to obtain uncalibrated measurements of electrical potentials. All experimental details are described in~\cite{Chowdhury2018}. %Although optical mapping can be challenging \textit{in vivo}, \textit{in vitro} experiments provides detailed insights into cellular level EP and thus arrhythmic mechanisms.

We investigated the effects of four drugs: HMR1556, nifedipine, lidocaine, and carbenoxolone (CBX). Their effects are summarised in Table~\ref{tab:drug_effects}. We assessed the effects of drugs on PINNs' estimates of model parameters. We identified drug-induced significant differences using a one-tailed two-sample $t$-test at a significance level of $0.05$.

\begin{table}[t]
\centering
\small % Reduce font size to small
\begin{tabular}{@{}lllll@{}}
\toprule
& \textbf{HMR1556} & \textbf{Nifedipine} & \textbf{Lidocaine} & \textbf{CBX} \\ \midrule

\makecell[l]{Site of\\ blockade} & \makecell[l]{Potassium \\channel \\($I$\textsubscript{K})} & \makecell[l]{Calcium\\ channel\\ ($I$\textsubscript{Ca})} & \makecell[l]{Sodium\\ channel\\ ($I$\textsubscript{Na})} & \makecell[l]{Intercellular\\ gap\\ junctions} \\
 \addlinespace
\makecell[l]{Effects on\\ model\\ parameters} &\makecell[l] {Increases\\ $\tau_{r}$} &\makecell[l] {Increases\\ $\tau_{si}$} &\makecell[l] {Increases\\ $\tau_{d}$} & \makecell[l]{No effect\\ on all $\tau$} \\
 \addlinespace
\makecell[l]{Effects\\ on AP} & \makecell[l]{Prolongs\\ duration} & \makecell[l]{Shortens\\ duration} & \makecell[l]{Prolongs\\ upstroke} & \makecell[l]{Decreases\\ conduction\\ velocity} \\
\bottomrule
\end{tabular}
\caption{Effects of anti-arrhythmic drugs~\cite{thomas2003hmr,godfraind2017discovery,bean1983lidocaine,kojodjojo2006effects}.}
\label{tab:drug_effects}
\end{table}

%They respectively reduce the ionic conductances of: the delayed rectifier potassium channel ($I$\textsubscript{Ks})~\cite{thomas2003hmr}, L-type calcium channel ($I$\textsubscript{CaL})~\cite{godfraind2017discovery}, sodium channel ($I$\textsubscript{Na})~\cite{bean1983lidocaine} and gap junctions~\cite{kojodjojo2006effects}. The first three drugs corresponds to an increase in the Fenton-Karma model parameters $\tau_r$, $\tau_{si}$, $\tau_{d}$, respectively. Carbenoxolone should not affect any of the $\tau$ parameters. HMR1556 prolongs the AP duration, nifedipine shortens it, and lidocaine slows down the AP upstroke. Carbenoxolone's main action is to slow down the AP conduction velocity. 

\begin{figure}[b]
    \centering
    \includegraphics[width=8.6cm]{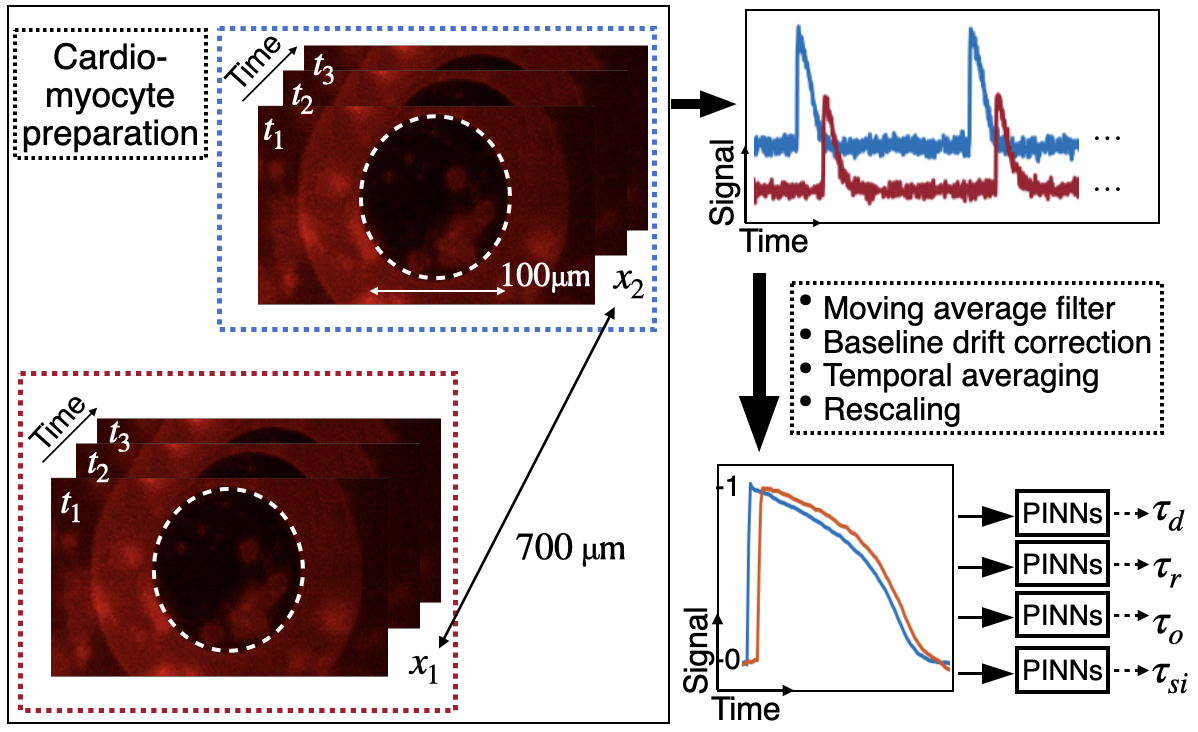}
    \caption{Data processing pipeline for optical mapping images.}
    \label{fig:OM_images}
\end{figure}

For each drug, we used four time series of optical mapping images. Two of these series were acquired under the action of the drug at half-maximal inhibitory concentration (IC\textsubscript{50}),  at two different preparation locations. The other pair was acquired in the absence of drugs (baseline conditions) also at two different locations. For each time series, we improved the signal-to-noise ratio of the data by:
\begin{itemize}
    \item spatially averaging the signal over a circular area of radius \num{50}\unit{\micro\metre} (indicated in Fig.~\ref{fig:OM_images});
    \item applying a temporal moving average filter to the signal except at the AP upstrokes;
    \item correcting temporal baseline drifts by subtracting each AP by the average of \num{10}\unit{\milli\second} before each upstroke;
    \item overlaying and averaging all APs across time;
    \item and scaling each averaged AP to the [0, 1] interval. 
\end{itemize}

For simplicity, we modelled the cardiomyocyte preparation as a \num{10}\unit{mm} cable with the measured APs placed at positions $x_1 = 2$\unit{mm} and $x_2 = 8$\unit{mm}. A diffusion coefficient of $D=0.1$ \unit{\milli\metre\per\milli\second} was used. These were then used as inputs for PINNs. This pipeline is illustrated in Fig.~\ref{fig:OM_images}.

% We manually selected two square regions of interest (ROIs) with a side of $2.3~\mu m$ and with their centres $0.7~mm$ apart, in the same image location for each time series. We spatially averaged the optical signal over each ROI to obtain a signal trace across time. From this signal, we manually selected two consecutive APs \hl{and normalised the signal to the [0, 1] interval for consistency with the Aliev-Panfilov model}. To improve the signal to noise ratio (SNR) of this trace, we applied a mean average filter twice, aligned and averaged the two APs over time to obtain a single higher-SNR AP. These pairs of post-processed APs were used as inputs to 1D EP-PINNs in inverse mode, with $b$ as the variable to be estimated. All model parameters were unchanged from those in Supplementary Table 1. We used EP-PINNs' architecture A with training scheme 2 to estimate $b$ 10 times for each setting. 116-200 points were used for training EP-PINNs and 29-50 to test it, as detailed in Supplementary Table 2.
%, except $D =  mm^2/TU$.} 

\subsection{PINNs Setup}

We used a fully-connected neural network with 4 layers of 32 neurons each, a $tanh$ activation, a Glorot Uniform initialisation, and Adam optimisation. There were two input variables, space and time position $x$ and $t$, and three output variables: $u$, $v$, and $w$. We trained the network for $100,000$ epochs with learning rate $5\times 10^{-5}$. These hyperparameters were selected empirically based on tests using \textit{in silico} data. The loss function consists of 8 terms:
%\begin{equation}
\begin{align}
    L &= L_{fu} + L_{fv} + L_{fw}+ L_\text{BC} + \nonumber \\
    &L_{\text{IC}, u} + L_{\text{IC}, v}+L_{\text{IC}, w}+ L_\text{data}.
\end{align}
%\end{equation}
$L_{fu} + L_{fv} + L_{fw}$ accounts for the agreement with the Fenton-Karma equations Eq.~\eqref{FK1}-\eqref{FK3}. $L_\text{BC}$ and $L_{\text{IC}, u} + L_{\text{IC}, v} + L_{\text{IC}, w}$ accounts for the no-flux Neumann boundary condition for membrane voltage $\frac{\partial u}{\partial \vec{n}} = 0$ and initial conditions for variables $u$, $v$, and $w$. $L_\text{data}$ accounts for the agreement with experimental measurements of $u$. We empirically determined that equal weights for all terms leads to the best network performance. 
\begin{figure}
    \centering
    \includegraphics[width=8.6cm]{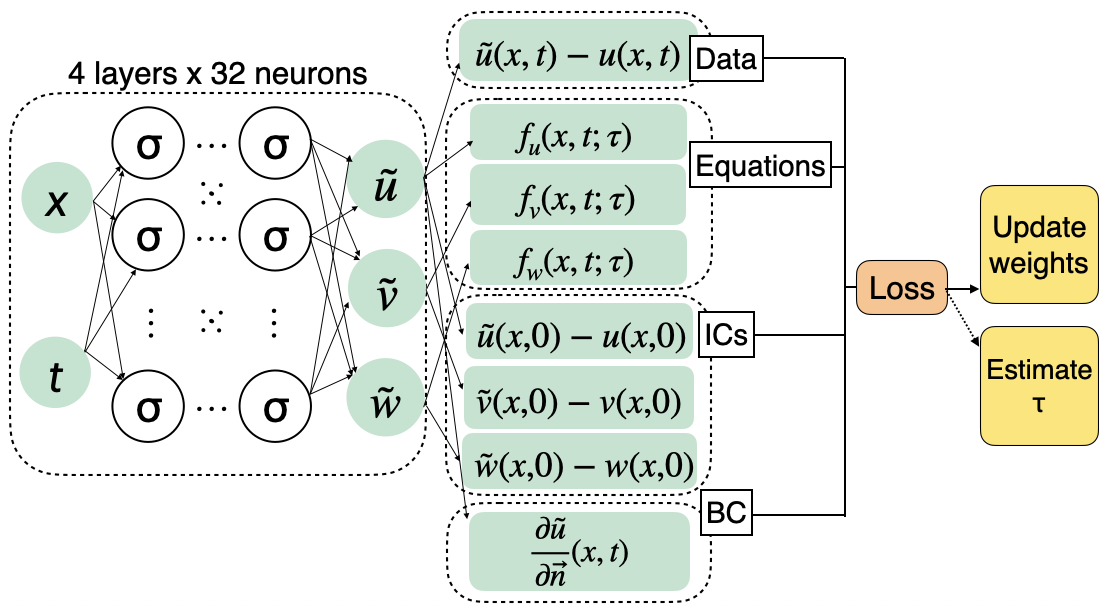}
    \caption{The PINNs architecture. Inputs $x$ and $t$ are the spatial and time points. The network optimises its weights by minimising the loss function $L$, to provide solutions for $u$, $v$, and $w$. It simultaneously estimates model parameter(s) $\tau$.}
    \label{fig:architecture}
\end{figure}
%Each term in the loss function is computed on different sets of points. 
% %\begin{equation}
% \begin{align}
% L = &\frac{1}{N_f} \sum_{j=1}^{N_f} (f_u({x_j},t_j)^2+ f_v({x_j},t_j)^2 + f_w({x_j},t_j)^2) + \nonumber \\
%     & \frac{1}{N_b} \sum_{k=1}^{N_b} (\frac{\partial u}{\partial \vec{n}}(x_k,t_k))^2 + \frac{1}{N_0} \sum_{l=1}^{N_0} (u({x_l},t_0) - u_0) ^2+ \nonumber \\
%     &\frac{1}{N} \sum_{i=1}^{N} (u({x_i},t_i)- {u_{GT}}_i)^2,
% \end{align}
% %\end{equation}
% where $(x_j, t_j)$ are the $N_f$ points within the domain of the equations. $(x_k, t_k)$ are the $N_b$ boundary points, $(x_l, t_0)$ the $N_0$ initial points, and $(x_i, t_i)$ the $N$ points where either the \textit{in vitro} or \textit{in silico} data are known.
For training, we used $60000$ points within the domain for the agreement with equations, $1000$ boundary points, and $99$ initial points for boundary and initial conditions. We split all available data into train and test sets at $9$:$1$ and calculate the root mean squared error (RMSE) on the test points. %We implemented an automatic reset when the losses at the first epoch exceed a predefined threshold to minimise convergence problems.

\subsection{Model Parameter Estimation using PINNs}
We used PINNs to estimate model parameters $\tau_d$, $\tau_r$, $\tau_o$, and $\tau_{si}$ in two modes: a) individually, with the other parameters fixed at literature values (see FK-CAF in~\cite{goodman2005membrane}), and b) estimating all four parameters simultaneously. We initialised the model parameter(s) by randomly drawing from a uniform distribution around the ground truth(s), where the distribution widths were based on uncertainties from preliminary results.

To estimate the uncertainty in parameter estimates, the network was randomly initialised to a different configuration five times and the standard deviation of the model parameter estimates over these runs was computed. 

\section{Results}
\label{sec:pagestyle}
\subsection{Parameter estimation with \textit{in silico} data}
\begin{figure}[htbp]
\centering
\includegraphics[width=8cm]{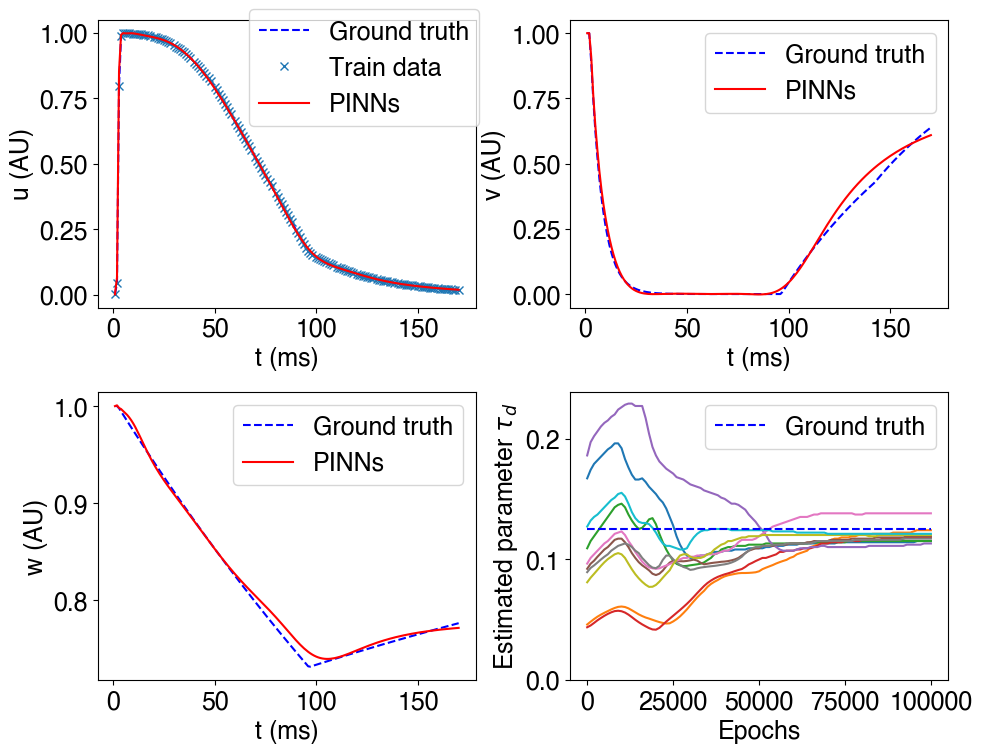}
    \caption{PINNs' predictions and \textit{in silico} ground truth data for transmembrane potential $u$ and gate variables $v$ and $w$ at \num{2}\unit{\milli\metre} on a 1D \num{10}\unit{\milli\metre} cable. Only data for $u$ were given here and at \num{8}\unit{\milli\metre}, and none for $v$ and $w$. Bottom right: 10 independent runs of parameter estimates of $\tau_d$ across epochs.}
    \label{fig:synthetic_uvw_plots}
\end{figure}
Using PINNs, we were able to accurately estimate the Fenton-Karma model parameters, and accurately reproduce the APs and time course of latent variables, as shown in Fig.~\ref{fig:synthetic_uvw_plots}. PINNs predictions for variables $u$, $v$, and $w$ are in good agreement with the synthetic ground truth data, even when only data for $u$ were given at two locations. Estimates of the variables were not affected by the number of model parameters predicted by PINNs: RMSE was $(4.6 \pm 2.0)\times 10^{-3}$ when estimating each parameter individually, and $(4.9 \pm 2.2)\times 10^{-3}$ when estimating all four simultaneously.

\begin{figure}[htbp]\centering
\includegraphics[width=6.6cm]{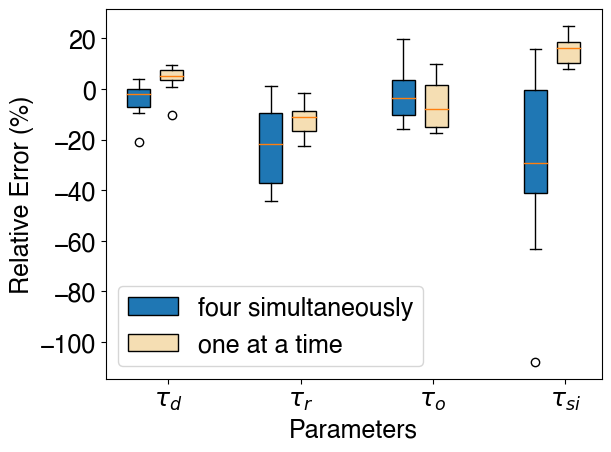}
    \caption{Relative errors of parameter estimates using \textit{in silico} data.}
    \label{fig:synthetic_RE}
\end{figure}

As shown in Fig.~\ref{fig:synthetic_RE}, when estimating each parameter separately, REs were within $15\%$. REs were generally larger when estimating four parameters simultaneously, as expected, but even in this regime did not exceed $30\%$. $\tau_{si}$ (inverse of $I$\textsubscript{Ca} conductance) was the least accurate in both modes, and $\tau_d$ (inverse of $I$\textsubscript{Na} conductance) the most accurate.

\subsection{Drug Effect characterisation using PINNs}
\begin{figure}[h!]
\centering
\includegraphics[width=8.6cm]{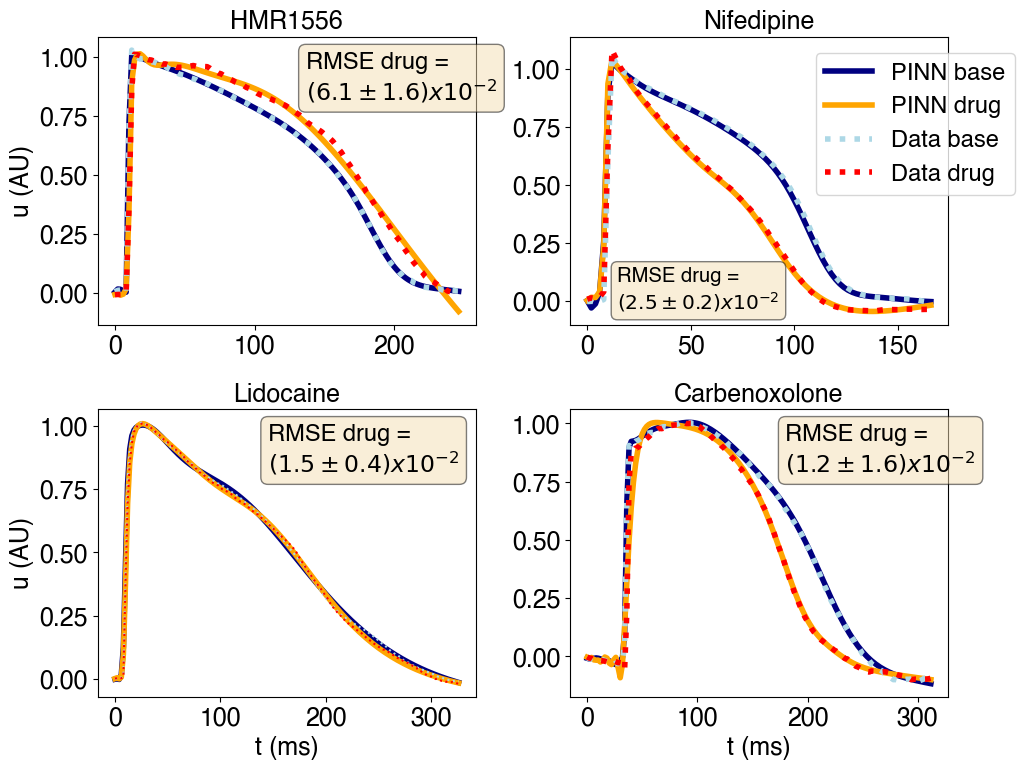}
    \caption{PINNs' predictions of the action potentials and \textit{in vitro} optical mapping data showed good agreement even when some model parameters were not given to the network. The RMSE for $u$ in baseline conditions was $(2.6 \pm 2.5)\times10^{-2}$.}
    \label{fig:in vitro APs}
\end{figure}

With experimental \textit{in vitro} data as inputs, PINNs were also able to accurately reproduce the experimental APs, with an average RMSE for $u$ within $7\times 10^{-2}$  (Fig.~\ref{fig:in vitro APs}). %For HMR1556, the RMSE for $u$ estimation was $(6.1\pm 1.6)\times 10^{-2}$; for nifedepine $(2.5 \pm 1.5)\times 10^{-3}$; for lidocaine $(1.5 \pm 3.6)\times 10^{-3}$; and for CBX $(1.2 \pm 1.6)\times 10^{-2}$. The RMSE for baseline $u$ is $(2.6 \pm 2.5)\times10^{-2}$. 

Our method remarkably identified the effects of drugs on each ionic channel by measuring significant increases in the $\tau$ parameters that correspond to the ionic channels each drug is known to modulate.
The relative changes in $\tau$ and the associated $p$-values are reported in Table~\ref{tab:drug_results}. We estimated each parameter separately for \textit{in vitro} data for stability.

\begin{table}[h!]
  \centering
  \small
  % Table for HMR1556 and Nifedipine
  \begin{tabular}{@{}lcc@{}}
    \toprule
    & HMR1556 & Nifedipine \\
    \midrule
    \(\Delta \tau_d\) & \(-3.4\pm0.3\%\) & \(-5.5\pm0.5\%\) \\
    \(p\)-value & 0.50 & 0.28 \\
    \addlinespace
    \(\Delta \tau_r\) & \cellcolor{green!25}\(31.8\pm2.7\%\) & \(-0.69\pm0.04\%\) \\
    \(p\)-value & \cellcolor{green!25} \(8.2\times10^{-5}\) & 0.85 \\
    \addlinespace
    \(\Delta \tau_o\) & \(37.1\pm14.5\%\) & \(9.5\pm1.4\%\) \\
    \(p\)-value & 0.065 & 0.24 \\
    \addlinespace
    \(\Delta \tau_{si}\) & \(-13.6\pm2.6\%\) & \cellcolor{green!25}\(80.9\pm21.6\%\) \\
    \(p\)-value & 0.19 & \cellcolor{green!25}0.024 \\
    \addlinespace
    Expected & Increases \(\tau_r\) & Increases \(\tau_{si}\) \\
    \addlinespace
    %RMSE for $u$ & $6.6\times 10^{-2} \pm 1.1\times 10^{-2}$ & $2.5\times 10^{-2} \pm 1.5\times 10^{-3}$ \\
    \bottomrule
  \end{tabular}

  \vspace{0.1em} % Space between tables

  % Table for Lidocaine and Carbenoxolone
  \begin{tabular}{@{}lcc@{}}
    \toprule
    & Lidocaine & CBX \\
    \midrule
    \(\Delta \tau_d\) & \cellcolor{green!25}\(8.6\pm0.5\%\) & \(-8.3\pm4.6\%\) \\
    \(p\)-value & \cellcolor{green!25}0.03 & 0.73 \\
    \addlinespace
    \(\Delta \tau_r\) & \(1.2\pm0.2\%\) & \(9.1\pm0.9\%\) \\
    \(p\)-value & 0.84 & 0.09 \\
    \addlinespace
    \(\Delta \tau_o\) & \(17.7\pm10.6\%\) & \(-8.1\pm6.2\%\) \\
    \(p\)-value & 0.58 & 0.81 \\
    \addlinespace
    \(\Delta \tau_{si}\) & \(-9.8\pm2.0\%\) & \(-2.2\pm0.2\%\) \\
    \(p\)-value & 0.22 & 0.56 \\
    \addlinespace
    Expected & Increases \(\tau_d\) & No change in \(\tau\) \\
    \bottomrule
  \end{tabular}
  \caption{Relative changes in the \(\tau\) parameters (inverse of channel conductance) between baseline and drug and associated $p$-values. Significant increases in $\tau$ are shown in green.}
  \label{tab:drug_results}
\end{table}

% \section{Type-style and fonts}
% \label{sec:typestyle}

% To achieve the best rendering both in the printed and digital proceedings, we
% strongly encourage you to use Times-Roman font.  In addition, this will give
% the proceedings a more uniform look.  Use a font that is no smaller than nine
% point type throughout the paper, including figure captions.

% In nine point type font, capital letters are 2 mm high.  If you use the
% smallest point size, there should be no more than 3.2 lines/cm (8 lines/inch)
% vertically.  This is a minimum spacing; 2.75 lines/cm (7 lines/inch) will make
% the paper much more readable.  Larger type sizes require correspondingly larger
% vertical spacing.  Please do not double-space your paper.  True-Type 1 fonts
% are preferred.

% The first paragraph in each section should not be indented, but all the
% following paragraphs within the section should be indented as these paragraphs
% demonstrate.

\section{Discussion}
\label{sec:majhead}
In this work, we developed a PINNs framework capable of accurate characterisation of anti-arrhythmic drug actions on EP parameters, using experimental \textit{in vitro} images. We coupled PINNs with the Fenton-Karma model, which distinguishes among currents of the three main ion species in the heart: Na$^+$, K$^+$, and Ca$^{2+}$. Starting with \textit{in silico} data, PINNs accurately estimated the model parameters corresponding to different ionic channel conductances. Then, with \textit{in vitro} optical mapping images, PINNs identified significant increases in model parameters corresponding to the expected drug-induced reduction of channel conductances. In both cases, PINNs gave excellent predictions for AP propagation, which demonstrated its capability to work well with sparse measurements, experimental noise and artefacts.

So far, studies have mostly focussed on demonstrating PINNs performance on \textit{in silico} data~\cite{sahli2020physics, Yazdani2020,Lu2021DeepXDE:Equations}. We showed that PINNs are also effective for model calibration with \textit{in vitro} data as inputs. Previous work had used PINNs to characterise drug effects with a simple model that does not consider ionic channels~\cite{herrero2022ep}. Our work shows that it is possible to use more sophisticated cardiac EP models to gain more detailed insights into drug mechanisms, which has important applications in the safety and efficacy testing of new drugs~\cite{zemzemi2013computational}.
% Address comparison with alternative approaches
Compared to traditional numerical methods, our approach has the advantages of being mesh-free, using automatic differentiation, and can be compactly implemented to simultaneously solve forward and inverse problems~\cite{Lu2021DeepXDE:Equations}.

% Address the disadvantages of approach
Limitations in our current framework include experiment-specific results, where evaluation on other sets of data would be valuable for robustness. In addition, the parameter estimates and convergence properties are sensitive to the initialisation. Future work may focus on exploring alternative optimisation schemes and loss function compositions. We also plan to extend this framework to the analysis of electrograms to bring PINNs' advantages closer to the clinical setting.

\section{Compliance with ethical standards}
\label{sec:ethics}
This is a computational study for which no ethical approval was required. The data used were acquired in~\cite{Chowdhury2018}, with the appropriate ethical approvals detailed therein: all procedures were conducted according to the standards set by the EU Directive 2010/63/EU and were approved by the Imperial College London Ethical Committee.

\section{Acknowledgments}
\label{sec:acknowledgments}

This work was supported by the  British Heart Foundation (RE/18/4/34215, RG/16/3/32175, PG/16/17/32069 and Centre of Research Excellence), the National Institute for Health Research (UK) Biomedical Research Centre and the Rosetrees Trust through the interdisciplinary award "Atrial Fibrillation: A Major Clinical Challenge".

% IEEE-ISBI supports the disclosure of financial support for the project
% as well as any financial and personal relationships of the author that
% could create even the appearance of bias in the published work. The
% authors must disclose any agency or individual that provided financial
% support for the work as well as any personal or financial or
% employment relationship between any author and the sources of
% financial support for the work.

% Other types of acknowledgements can also be listed in this section.

% Reporting on real or potential conflicts of interests, or the absence
% thereof, is required in the paper. Authors are responsible for
% correctness of the statements provided in the manuscript. Examples of
% appropriate statements include:
% \begin{itemize}
%   \item ``No funding was received for conducting this study. The
%     authors have no relevant financial or non-financial interests to
%     disclose.'' 
%   \item ``This work was supported by […] (Grant numbers) and
%     […]. Author X has served on advisory boards for Company Y.'' 
%   \item ``Author X is partially funded by Y. Author Z is a Founder and
%     Director for Company C.''
% \end{itemize}

\bibliographystyle{IEEEbib}
\bibliography{strings,references}

\begin{thebibliography}{10}

\bibitem{Raissi2019}
M.~Raissi, P.~Perdikaris, and G.~E. Karniadakis,
\newblock ``{Physics-informed neural networks: A deep learning framework for solving forward and inverse problems involving nonlinear partial differential equations},''
\newblock {\em Journal of Computational Physics}, vol. 378, pp. 686--707, 2 2019.

\bibitem{sahli2020physics}
Francisco Sahli~Costabal, Yibo Yang, Paris Perdikaris, Daniel~E Hurtado, and Ellen Kuhl,
\newblock ``Physics-informed neural networks for cardiac activation mapping,''
\newblock {\em Frontiers in Physics}, vol. 8, pp. 42, 2020.

\bibitem{grandits2021learning}
Thomas Grandits, Simone Pezzuto, Francisco~Sahli Costabal, Paris Perdikaris, Thomas Pock, Gernot Plank, and Rolf Krause,
\newblock ``Learning atrial fiber orientations and conductivity tensors from intracardiac maps using physics-informed neural networks,''
\newblock in {\em International Conference on Functional Imaging and Modeling of the Heart}. Springer, 2021, pp. 650--658.

\bibitem{herrero2022ep}
Clara Herrero~Martin, Alon Oved, Rasheda~A Chowdhury, Elisabeth Ullmann, Nicholas~S Peters, Anil~A Bharath, and Marta Varela,
\newblock ``Ep-pinns: Cardiac electrophysiology characterisation using physics-informed neural networks,''
\newblock {\em Frontiers in Cardiovascular Medicine}, vol. 8, pp. 768419, 2022.

\bibitem{aliev1996simple}
Rubin~R Aliev and Alexander~V Panfilov,
\newblock ``A simple two-variable model of cardiac excitation,''
\newblock {\em Chaos, Solitons \& Fractals}, vol. 7, no. 3, pp. 293--301, 1996.

\bibitem{fenton1998vortex}
Flavio Fenton and Alain Karma,
\newblock ``Vortex dynamics in three-dimensional continuous myocardium with fiber rotation: Filament instability and fibrillation,''
\newblock {\em Chaos: An Interdisciplinary Journal of Nonlinear Science}, vol. 8, no. 1, pp. 20--47, 1998.

\bibitem{goodman2005membrane}
Amy~M Goodman, Robert~A Oliver, Craig~S Henriquez, and Patrick~D Wolf,
\newblock ``A membrane model of electrically remodelled atrial myocardium derived from in vivo measurements,''
\newblock {\em EP Europace}, vol. 7, no. s2, pp. S135--S145, 2005.

\bibitem{Chowdhury2018}
Rasheda~A Chowdhury, Konstantinos~N. Tzortzis, Emmanuel Dupont, Shaun Selvadurai, Filippo Perbellini, Chris~D. Cantwell, Fu~Siong Ng, Andre~R. Simon, Cesare~M. Terracciano, and Nicholas~S. Peters,
\newblock ``{Concurrent micro-to macro-cardiac electrophysiology in myocyte cultures and human heart slices},''
\newblock {\em Scientific Reports}, vol. 8, no. 1, pp. 1--13, 12 2018.

\bibitem{thomas2003hmr}
George~P Thomas, Uwe Gerlach, and Charles Antzelevitch,
\newblock ``Hmr 1556, a potent and selective blocker of slowly activating delayed rectifier potassium current,''
\newblock {\em Journal of cardiovascular pharmacology}, vol. 41, no. 1, pp. 140--147, 2003.

\bibitem{godfraind2017discovery}
Th{\'e}ophile Godfraind,
\newblock ``Discovery and development of calcium channel blockers,''
\newblock {\em Frontiers in pharmacology}, vol. 8, pp. 286, 2017.

\bibitem{bean1983lidocaine}
BRUCE~P Bean, CHARLES~J Cohen, and RICHARD~W Tsien,
\newblock ``Lidocaine block of cardiac sodium channels,''
\newblock {\em The Journal of general physiology}, vol. 81, no. 5, pp. 613, 1983.

\bibitem{kojodjojo2006effects}
Pipin Kojodjojo, Prapa Kanagaratnam, Oliver~R Segal, Wajid Hussain, and Nicholas~S Peters,
\newblock ``The effects of carbenoxolone on human myocardial conduction: a tool to investigate the role of gap junctional uncoupling in human arrhythmogenesis,''
\newblock {\em Journal of the American College of Cardiology}, vol. 48, no. 6, pp. 1242--1249, 2006.

\bibitem{Yazdani2020}
Alireza Yazdani, Lu~Lu, Maziar Raissi, and George~Em Karniadakis,
\newblock ``{Systems biology informed deep learning for inferring parameters and hidden dynamics},''
\newblock {\em PLoS Computational Biology}, vol. 16, no. 11, pp. 1--19, 2020.

\bibitem{Lu2021DeepXDE:Equations}
Lu~Lu, Xuhui Meng, Zhiping Mao, and G.~E. Karniadakis,
\newblock ``{DeepXDE: A deep learning library for solving differential equations},''
\newblock {\em SIAM Review}, vol. 63, no. 1, pp. 208--228, 2 2021.

\bibitem{zemzemi2013computational}
Nejib Zemzemi, Miguel~O Bernabeu, Javier Saiz, Jonathan Cooper, Pras Pathmanathan, Gary~R Mirams, Joe Pitt-Francis, and Blanca Rodriguez,
\newblock ``Computational assessment of drug-induced effects on the electrocardiogram: from ion channel to body surface potentials,''
\newblock {\em British journal of pharmacology}, vol. 168, no. 3, pp. 718--733, 2013.

\end{thebibliography}

\end{document}